\documentclass[manuscript,nonacm]{acmart}
\setcopyright{none}
\renewcommand\footnotetextcopyrightpermission[1]{}
\usepackage{subcaption}
\usepackage{makecell}
\AtBeginDocument{%
  }

\setcopyright{acmlicensed}
\copyrightyear{2025}
\acmYear{2025}
\acmDOI{XXXXXXX.XXXXXXX}
\acmConference[RISC-V HPC Asia '25]{International workshop on RISC-V for HPC at HPC Asia 2025}{February 19, 2025}{Hsinchu, Taiwan}
\acmISBN{978-1-4503-XXXX-X/2018/06}

\begin{document}

\title{Investigations of multi-socket high core count RISC-V for HPC workloads}

\author{Nick Brown}
\email{n.brown@epcc.ed.ac.uk}
\orcid{1234-5678-9012}
\author{Christopher Day}
\affiliation{%
  \institution{EPCC at the University of Edinburgh}
  \city{Edinburgh}  
  \country{UK}
}

\renewcommand{\shortauthors}{Brown et al.}

\begin{abstract}
  Whilst RISC-V has become popular in fields such as embedded computing, it is yet to find mainstream success in High Performance Computing (HPC). However, the 64-core RISC-V Sophon SG2042 is a potential game changer as it provides a commodity available CPU with much higher core count than existing technologies. In this work we benchmark the SG2042 CPU hosted in an experimental, dual-socket, system to explore the performance properties of the CPU when running a common HPC benchmark suite across sockets. Earlier benchmarks found that, on the Milk-V Pioneer workstation, whilst the SG2042 performs well for compute bound codes, it struggles when pressure is placed on the memory subsystem. The performance results reported here confirm that, even on a different system, these memory performance limitations are still present and hence inherent in the CPU. However, a multi-socket configuration does enable the CPU to scale to a larger number of threads which, in the main, delivers an improvement in performance and-so this is a realistic system configuration for the HPC community.
\end{abstract}

\ccsdesc[500]{Computer systems organization~Multicore architectures}
\ccsdesc[300]{Computing methodologies~Concurrent computing methodologies}

\keywords{RISC-V, High Performance Computing (HPC),  Sophon SG2042, NAS Parallel Benchmark suite (NPB)}

\maketitle

\section{Introduction}

RISC-V is an open Instruction Set Architecture (ISA) that has enjoyed wide popularity since it was first released over a decade ago. Indeed, in late 2024 Nvidia announced that there were multiple RISC-V cores in each of their GPUs, however, for all these successes RISC-V has yet to gain widespread traction as a main-line technology in High Performance Computing (HPC). As the HPC community moves further into the exascale era, and there is increased focus on sustainability, the community must tackle the challenge around how to deliver both increased performance and greater energy efficiency. Consequently, there are opportunities for technologies such as RISC-V, where vendors can more easily integrate specialisations, to fill an important role in HPC. 

A common criticism from the HPC community about RISC-V is that there is a lack of real-world high performance hardware available to build systems around. A major step forwards was therefore in summer 2023 when Sophon announced the SG2042, which is the first high core count commodity available RISC-V CPU. Containing 64-cores, the SG2042 RISC-V CPU is aimed at high performance workloads and consequently this is a far more serious proposition for the HPC community than previous RISC-V hardware. Previous work \cite{brown2023risc} found that whilst each core of the SG2042 was a considerable improvement beyond other existing RISC-V hardware, the CPU struggled to match the performance of x86-based CPUs that are commonplace in HPC. However, from this previous work it was not possible to understand whether the issues that were observed were inherent to CPU itself or the rest of the system, the Milk-V Pioneer. The Pioneer is a designed as a workstation and-so there could be limitations more generally around the memory subsystem that then impact the performance observed when running codes on this system. In this paper we build on previous benchmarking work to revisit previously run benchmarks on a bespoke dual-socket SG2042 based system developed by E4 Computer Engineering SPA. Comparing against previous published results, we compare and contrast the insights gained previously against the observed performance on the E4 system. This paper is structured as follows; after describing the background to this work in Section \ref{sec:bg}, we then explore the performance across the two sockets on two different nodes in Section \ref{{sec:socket-variance}} of the test system to understand how consistent performance is across socket. Based upon these insights we then select the best performing configuration and compare this against AMD EPYC, Intel Skylake and Marvel ThunderX2 CPUs in Section \ref{sec:other-hw}. Lastly we draw conclusions to this work in Section \ref{sec:conc}.

\section{Background}
\label{sec:bg}


The Sophon SG2042 CPU is a 64-core processor operating at 2GHz, where the XuanTie C920 cores are organized into clusters of four. Each 64-bit C920 core, developed by T-Head, adopts an out-of-order, multiple-issue superscalar pipeline design \cite{c906}. The RV64GCV instruction set is implemented by the C920, and support for version 0.7.1 of the vectorization standard extension (RVV v0.7.1) \cite{c920-nofp64}, with a vector width of 128 bits, is provided. Each C920 core contains 64KB of L1 instruction (I) and data (D) cache, 1MB of L2 cache shared by the four-core cluster, and 64MB of L3 system cache shared across all cores in the package. The Sophon SG2042 also features four DDR4-3200 memory controllers and 32 lanes of PCI-E Gen4. 

Because the C920 core only supports RVV v0.7.1, which is not compatible with the mainline GCC or LLVM, T-Head released their own fork of the GNU compiler (XuanTie GCC). This is specifically optimized for their family of processor cores and supporting RVV v0.7.1. Prior work \cite{lee2023backporting} developed a tool that would backport RVV v1.0 to RVV 0.7.1, enabling a wider range of compilers such as LLVM to vectorise for the CPU. \cite{brown2023risc} undertook the first benchmarking of the Sophon SG2042 and the authors leveraged the RAJAPerf suite to compare and contrast against x86 based CPUs. However, the RAJAPerf suite comprises a large number of kernels and the results were such that whilst it was possible to generally observe that the Sophon SG2042 fell short of the performance delivered by these other CPUs, it was difficult to identify specific patterns and results behind the performance that was observed. 

Ultimately, an important question to answer was where the strengths and weaknesses of the SG2042 lie, and as such a follow on paper \cite{brown2025performance} leveraged the NAS Parallel Benchmark (NPB) suite \cite{benchmarks2006parallel} to undertake more targetted benchmarking of the CPU. By first performance profiling each targetted benchmark, the authors were able to demonstrate that, core for core, computationally bound workloads suited this RISC-V CPU rather well. Indeed the SG2042 performed comparatively to a Marvell ThunderX2 for compute based benchmarks. Moreover, whilst x86-based Intel Xeon Platinum (Skylake) and AMD Rome CPUs outperformed the RISC-V CPU, the large number of cores contained within the SG2042 meant that for these sorts of workloads it outperformed x86 CPUs with fewer cores, including the 26-core Intel Xeon Platinum.

However, when running benchmarks that were memory bound, either bandwidth or latency, the performance of the SG2042 was significantly below that of the Marvell ThunderX2 or x86 CPUs. It was observed in \cite{brown2025performance} that this was also a limitation beyond specific benchmarks because, when exploring performance of the NPB pseudo application benchmarks, the SG2042 delivered consistently lower performance than the other CPUs under test due to the pressure on the memory subsystem.

In \cite{diehl2024preparing} the authors explored running an astrophysics application on the SG2042 and found that the CPU delivered impressive performance compared to an A64FX. Furthermore, there were clear benefits in enabling vectorisation for this code compared to scalar runs. This work demonstrated that there is potential for such HPC applications to be executed on RISC-V CPUs, and specifically the SG2042.

However, the tests conducted in \cite{brown2025performance} and the performance numbers reported for the astrophysics application in \cite{diehl2024preparing} ran on a Sophon SG2042 which was as part of the Milk-V Pioneer. This is a single socket system that was designed for general workstation style tasks rather than HPC workloads. An important question is therefore whether the observed performance characteristics were due to the host system, rather than the CPU itself, and it is therefore instructive to explore the performance of the CPU within different host systems to gain more clarity. 

E4 Computer Engineering SPA have developed an experimental, dual socket SG2042 based system, as part of their RISC-V laboratory. Comprising a specially designed motherboard that mounts two Sophon SG2042 CPUs, along with 256GB of DDR memory, to the best of our knowledge this is the first multi-socket RISC-V high core count system based around a commodity CPU. It is therefore interesting to revisit these NPB benchmarks on the E4 system, in a multi-socket configuration, to better understand the performance properties and limitations of the SG2042 RISC-V CPU for high performance workloads. 

\subsection{NAS Parallel Benchmarks (NPB) suite}
\label{sec:npb}
NPB was developed in the 1990s by NASA's Advanced Supercomputing (NAS) division to better characterise the performance of their workloads on HPC systems. Focussing on Computational Fluid Dynamics (CFD) applications, the suite comprises individual kernel and pseudo application benchmarks. Whilst \cite{brown2025performance} explored both these kernels and pseudo applications on the Sophon SG2042, it was the kernels that provided the majority of the insight around the CPU's performance behaviour and-so these are the area of focus in this paper.

\begin{table}[htb]
    \centering
    \caption{Summary of memory behaviour for NPB benchmark kernels on a Xeon Platinum 8170 from \cite{brown2025performance}}
    \label{tab:npb_description}
    \begin{tabular}{|c|c|c|c|}
    \hline
      \textbf{Benchmark}   & \makecell{\textbf{Clock ticks} \\ \textbf{cache stall}} & \makecell{\textbf{Clock ticks} \\ \textbf{DDR stall}} & \makecell{\textbf{Time DDR} \\ \textbf{bandwidth bound}} \\
     \hline
	  \textbf{Integer Sort (IS)} & 35\% & 0\% & 16\% \\        
        \textbf{Embarrassingly Parallel (EP)} & 11\% & 0\% & 0\% \\
	  \textbf{Multi Grid (MG)} & 34\% & 20\% & 88\% \\        
        \textbf{Conjugate Gradient (CG)} & 19\% & 18\% & 0\% \\
        \textbf{Fast Fourier Transform (FT)} & 13\% & 9\% & 18\% \\        
    \hline
    \end{tabular}
\end{table}

In order to provide context to the performance results observed on the SG2042, the authors of \cite{brown2025performance} first profiled the NPB kernels on an x86 (Xeon Platinum 8170) system. This information is recreated in Table \ref{tab:npb_description}, where in this paper we explore the following kernels:

\begin{itemize}
    \item \textbf{Integer Sort (IS)}: is designed to test indirect, random, memory accesses and in Table \ref{tab:npb_description} it can be seen stalls the x86 CPU significantly due to cache misses. The indirect nature of the memory accesses means that the location of data retrieval is unpredictable and the kernel is memory latency bound.
    \item \textbf{Embarrassingly Parallel (EP)}: is compute bound and from Table \ref{tab:npb_description} it can be seen that, in comparison with IS, there are far fewer cycles stalled on memory access, and no time spent with high DDR bandwidth utilisation. This benchmark is designed to test the compute performance of hardware.   
    \item \textbf{Multi Grid (MG)}: is memory bound and it can be seen in Table \ref{tab:npb_description} that there is considerable time stalled on cache and main memory accesses, in addition to a significant amount of time where DDR is under high utilisation.
    \item \textbf{Conjugate Gradient (CG)}: requires irregular memory accesses and involves nearest neighbour communication which, in a shared memory system, will be undertaken via memory. This benchmark is therefore interesting because the nearest neighbour communication results in additional overhead.
    \item \textbf{Fourier Transform (FT)}: tests all-to-all communication between neighbours as part of the transposition. This is a notoriously difficult communication pattern to optimise for and can result in significant memory traffic and overhead.
\end{itemize}

All the NBP benchmarks are configured using a variety of problem sizes known as classes and in this paper we use class C throughout our experiments. There are numerous implementations of NBP to explore different technologies for the parallelisation, and in this paper we run the OpenMP versions without modification. All codes are compiled with optimization level three, each thread is mapped to a physical core, and the reported results are averaged over five runs. Each benchmark run reported in this paper was executed on the machine that was being used at the time exclusively for that run. When compiling for the SG2042 we use GNU Fortran (GCC) version 13.2.1 which is the default version on the E4 system.

\section{Performance variances across sockets and nodes}
\label{sec:socket-variance}
\begin{figure}
\begin{subfigure}{.5\textwidth}
  \centering
  \includegraphics[width=0.98\linewidth]{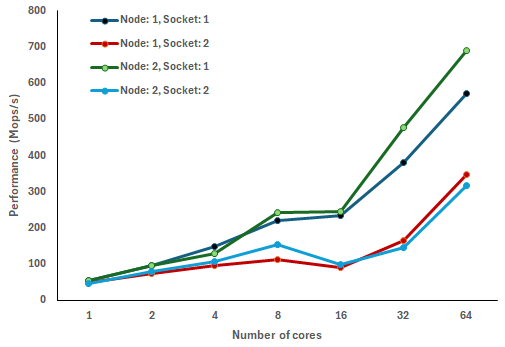}
  \caption{Integer Sort (IS)}
  \label{fig:var_is}
\end{subfigure}%
\begin{subfigure}{.5\textwidth}
  \centering
  \includegraphics[width=0.98\linewidth]{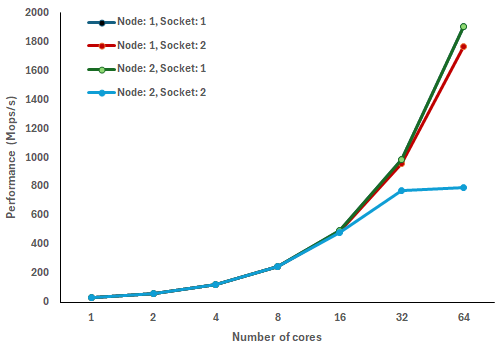}
  \caption{Embarrassingly Parallel (EP)}
  \label{fig:var_ep}
\end{subfigure}

\bigskip

\begin{subfigure}{.5\textwidth}
  \centering
  \includegraphics[width=0.98\linewidth]{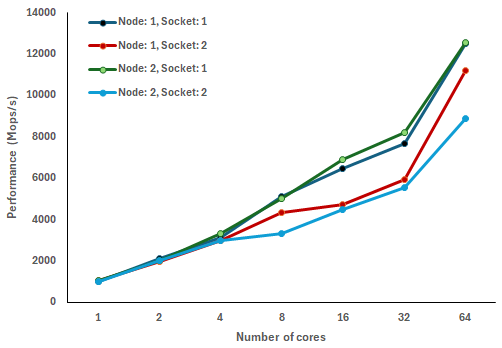}
  \caption{Multi Grid (MG)}
  \label{fig:var_mg}
\end{subfigure}%
\begin{subfigure}{.5\textwidth}
  \centering
  \includegraphics[width=0.98\linewidth]{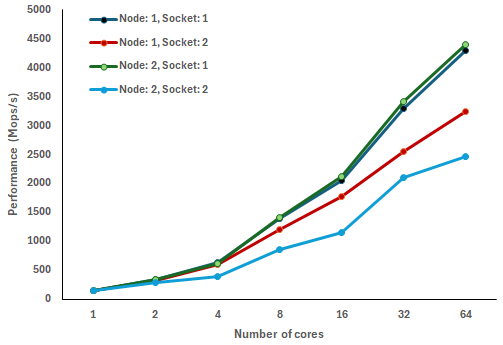}
  \caption{Conjugate Gradient (CG)}
  \label{fig:var_cg}
\end{subfigure}

\bigskip

\begin{subfigure}{.5\textwidth}
  \centering
  \includegraphics[width=0.98\linewidth]{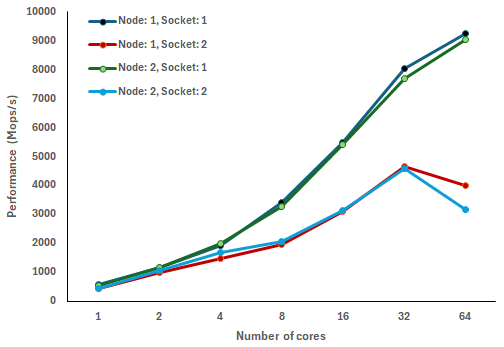}
  \caption{Fourier Transform (FT)}
  \label{fig:var_ft}
\end{subfigure}%

\caption{Performance (higher is better) when scaling number of threads within a socket. Experiments explore running sockets one and two of both nodes}
\label{fig:variance}
\end{figure}

The E4 system comprises two, dual-socket, nodes, and-so it is instructive to explore benchmark runs on different sockets and nodes to understand how consistent the resulting performance is. Figure \ref{fig:variance} illustrates, for each benchmark kernel, the performance (higher is better) as we scale the number of threads using OpenMP within a socket. For each benchmark run, performance numbers are reported on each socket for both nodes. 

From Figure \ref{fig:variance} it can be observed that there is a difference between sockets one and two in a node, where socket one outperforms socket two consistently across the benchmarks. The only exception to this is the EP benchmark, which tests compute performance, where sockets one and two on node one deliver very similar performance (and the lines are overlain in Figure \ref{fig:var_ep}), whereas there is a noteworthy difference between sockets one and two on node two for this same benchmark. The fact that socket one outperforms socket two for the memory bound benchmark kernels suggests that memory performance between socket two and DRAM is lower than socket one.

Both nodes in the E4 system are identical, so performance differences between nodes are interesting to observe. In the main, performance properties between the nodes is similar however as mentioned there is a significant difference for socket two when running the EP benchmark. It should be noted that this is an experimental system, and the first to leverage dual-socket SG2042 CPUs, therefore potentially there are opportunities for tuning that can address these variances and our hope is that these lessons learnt will assist the team.

\section{Comparison of dual-socket Sophon SG2042 against other architectures}
\label{sec:other-hw}
In this section we compare thread scaling on the Sophon SG2042 on up to 128 cores against CPUs of other architectures that are commonplace in HPC. Following the experiments run in \cite{brown2025performance} we use the same comparison CPUs, and Table \ref{tab:other_arch} summarises these. The AMD EPYC is the Rome series with the Zen-2 micro architecture. This CPU is contained within ARCHER2, which is a Cray EX and the UK national supercomputer and each node contains 256GB of DRAM. The EPYC is interesting here because, like the SG2042, it also contains 64 cores, and each node of ARCHER2 contains two sockets. Therefore, like for like, the CPU arrangement in a node of ARCHER2 is very similar to that of the RISC-V test machine. Furthermore, both the SG2042 and AMD EPYC CPUs contain four NUMA regions, each with 16 cores. There are some differences however, for instance the AMD EPYC has eight memory controllers and channels compared to four in the SG2042. Furthermore, the EPYC provides 256-bit wide vectorisation (via AVX2) compared to 128-bit wide in the SG2042, and the AMD CPU can process two AVX2 instructions per cycle. Each core in the AMD EPYC contains half the L1 cache (32KB of I and D), and double the L2 cache (512 KB per core, vs 1MB shared between 4 cores in the SG2042). The L3 cache (16MB shared between four cores) is four times greater than that in the SG2042. Codes are compiled using GCC version 11.2 on ARCHER2 and Simultaneous Multithreading (SMT) is disabled.

\begin{table}[]
    \centering
    \caption{Summary of CPUs that are benchmarked in this section from \cite{brown2025performance}}
    \label{tab:other_arch}
    \begin{tabular}{|c|c|c|c|c|c|}
    \hline
      \textbf{CPU} & \textbf{ISA} & \textbf{Part} & \textbf{Base clock} & \makecell{\textbf{Number} \\ \textbf{of cores}} & \textbf{Vector} \\
     \hline
	 AMD EPYC & x86-64 & EPYC 7742 & 2.25GHz & 64 & AVX2\\
      Intel Skylake & x86-64 & Xeon Platinum 8170 & 2.1 GHz & 26 & AVX512\\      
      Marvell ThunderX2 & ARMv8.1 & CN9980 & 2 GHz & 32 & NEON\\
      Sophon SG2042 & RV64GCV & SG2042 & 2 GHz & 64 & RVV v0.7.1\\
    \hline
    \end{tabular}
\end{table}

In order to explore the performance benefits of 512-bit vectorisation we also compare against an Intel Xeon Platinum 8170 (Skylake). There are 26-cores in this CPU and, in addition to providing two FPUs supporting AVX-512, each core contains half (32KB) the amount of L1 cache as the SG2042, four times the of L2 cache (1MB per core, whereas in the SG2042 the 1MB is shared between four cores), and 1.375MB of L3 cache which is a larger amount of L3 cache per core compared to the SG2042, but fewer cores in the Skylake means that in total the Intel CPU contains around two thirds of that in the SG2042. The machine we run on has 192GB of DDR4 memory, and we use GCC version 8.4.

Lastly, we compare against the CN9980 Marvell ThunderX2 containing 32 cores that implement the ARMv8.1 (AArch64) ISA via the Vulcan micro architecture. Each core contains half the L1 cache of the SG2042, and the same amount of L2 cache. There is a total of 32MB L3 cache, 1MB per core, which is the same per-core L3 cache as the SG2042 but half the overall amount due to the lower number of cores. The ThunderX2 provides 128 bit wide vectoristion via NEON which matches the width of the C920 core in the SG2042. However, the ThunderX2 contains two FPUs per core compared to only one in the C920. The machine contains 128GB of DDR per node, we use GCC version 9.2 and SMT is also disabled in our runs. 

A major difference between the experiments undertaken in this section and those of \cite{brown2025performance} is that previous work only ran over a single CPU, whereas in this work for the Sophon SG2042, AMD EPYC and Marvel ThunderX2 we scale across two sockets up to 128, 128 and 64 cores respectively. This is important because it is more realistic of how HPC users would run their OpenMP codes in practice, and also interesting for benchmarks that communicate because we can observe the difference that result from inter-socket runs.

\begin{figure}[htb]
\begin{subfigure}{.5\textwidth}
  \centering
  \includegraphics[width=0.98\linewidth]{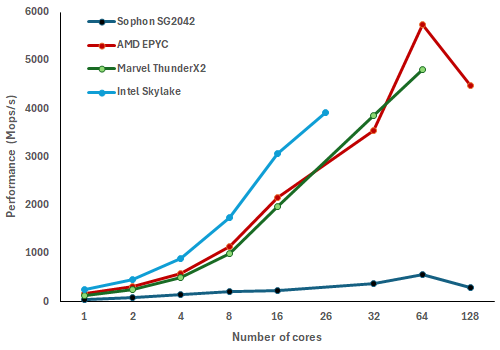}
  \caption{Integer Sort (IS)}
  \label{fig:total_is}
\end{subfigure}%
\begin{subfigure}{.5\textwidth}
  \centering
  \includegraphics[width=0.98\linewidth]{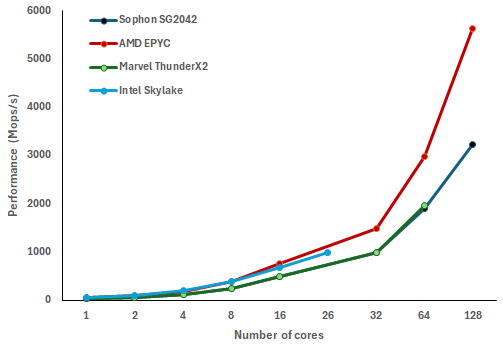}
  \caption{Embarrassingly Parallel (EP)}
  \label{fig:total_ep}
\end{subfigure}

\bigskip

\begin{subfigure}{.5\textwidth}
  \centering
  \includegraphics[width=0.98\linewidth]{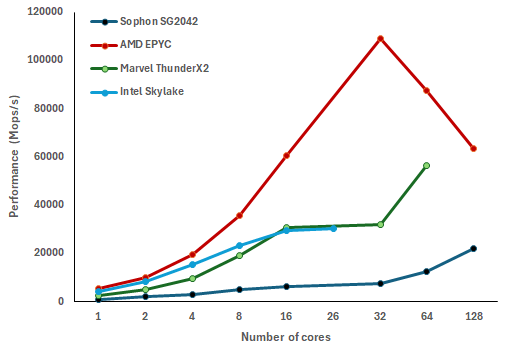}
  \caption{Multi Grid (MG)}
  \label{fig:total_mg}
\end{subfigure}%
\begin{subfigure}{.5\textwidth}
  \centering
  \includegraphics[width=0.98\linewidth]{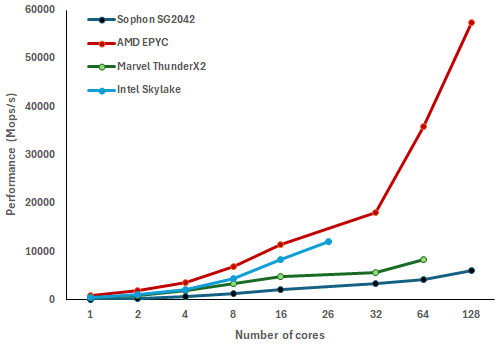}
  \caption{Conjugate Gradient (CG)}
  \label{fig:total_cg}
\end{subfigure}

\bigskip

\begin{subfigure}{.5\textwidth}
  \centering
  \includegraphics[width=0.98\linewidth]{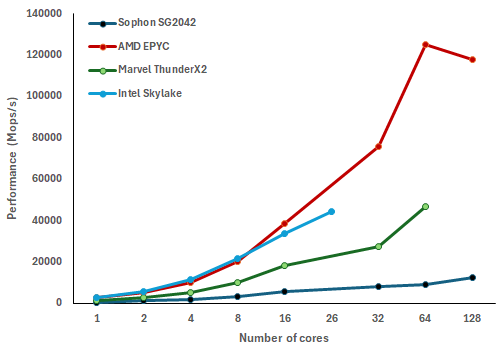}
  \caption{Fourier Transform (FT)}
  \label{fig:total_ft}
\end{subfigure}%

\caption{Performance (higher is better) when scaling number of threads comparing different architectures}
\label{fig:total}
\end{figure}

Figure \ref{fig:total} reports performance results for the five benchmark kernels across our four CPUs of interest. It can be seen that the Sophon SG2042 performs poorly against these other three CPUs for the IS benchmark, where the Skylake performs the best up until its 26 cores are exhausted, and the AMD EPYC and Marvel ThunderX2 deliver comparable performance. Notably, the second socket enables the Marvell ThunderX2 to continue scaling, whereas running across the second socket for the AMD EPYC and Sophon SG2042  decreases performance. The theory in \cite{brown2025performance} was that the poor performance of the SG2042 could be due to the design of cache hierarchy and limited memory bandwidth, however when running across sockets it can be observed there might also be an overhead between NUMA regions sitting on different sockets. The indirect memory accesses means that cores will inevitably need to unpredictably access memory in the NUMA region of another socket, and the AMD EPYC and Sophon SG2042 results in Figure \ref{fig:total_is} suggest that this results in additional overhead, over and above, accesses between NUMA regions within a socket.

In line with the results in \cite{brown2025performance}, the Sophon SG2042 delivers impressive performance for the EP benchmark in Figure \ref{fig:total_ep} which is compute bound. The AMD EPYC, Marvel ThunderX2 and Sophon SG2042 all scale well for this benchmark across the second socket, and the higher core count of the SG2042 delivers a noteworthy performance improvement over and above both the Intel Skylake and Marvel ThunderX2 which have a fewer total number of cores. As per \cite{brown2025performance}, the Marvel ThunderX2 and Sophon SG2042 both deliver very similar performance and this continues at 64 cores where the ThunderX2 is running across both sockets. This very similar performance is likely, in part, due to them both having 128-bit wide vector support. The AMD EPYC CPU has double the vector width which, in part, explains the performance gap between it and the Marvel ThunderX2 and Sophon SG2042 CPUs. 

The Multi Grid (MG) benchmark is memory bandwidth bound, and in Figure \ref{fig:total_mg} it can be seen that whilst the AMD EPYC provides considerably better performance, this performance decreases sharply at 64 and 128 cores. The Skylake and ThunderX2 deliver similar performance and both plateau at 16 cores, but the ThunderX2 then scales again as the benchmark is run across both sockets. The Sophon SG2042 lags the other CPUs, however at both 64 and 128 cores there is a performance uplift. The theory in \cite{brown2025performance} was that the memory controllers in the SG2042 are considerably less advanced than the other CPUs considered here. Although details on the memory controllers for th SG2042 are difficult to come by, the fact that the SG2042 then scales well across the second socket, effectively bringing an additional four memory controllers into play, tends to backup this hypothesis.

The Sophon SG2042 is outperformed by the other CPUs for both the CG and FT benchmarks, where it is beneficial scale to a second socket for both the Sophon SG2042 and the Marvel ThunderX2. However, the AMD EPYC decreases in performance for the FT benchmark when running on 128 cores across both sockets. Given that the CG benchmark only contains nearest neighbour communications, but the FT benchmark involves all-to-all communications, this further strengthens the hypothesis that on the AMD EPYC there is additional overhead in going between NUMA regions that sit on separate CPUs.

\section{Conclusions}
\label{sec:conc}
In this paper we have explored the performance of a dual-socket high core count RISC-V system for HPC workloads. Containing two 64-core SG2042 CPUs, the E4 Computer Engineering SPA system is interesting to explore firstly because it provides a different host to the Milk-V Pioneer workstation that was used previously to benchmark the SG2042, and secondly because dual-socket systems are more representative of nodes in HPC machines. This first aspect helps us to understand whether the performance properties of the SG2042 that were previously observed are inherent to the CPU or also indicative of the wider system. The second aspect enables us to explore configurations that are more realistic around how the HPC community would likely run codes on the SG2042 in production.

Leveraging NASA's NAS Parallel Benchmark suite (NPB), we demonstrated that the memory bound nature of the SG2042 observed in \cite{brown2025performance} is also seen on the E4 system and hence inherent in the design of the CPU itself. Scaling across a second socket resulted in improved performance on SG2042 for all but the IS benchmark and this demonstrated that, in the main, leveraging the SG2042 in a multi-socket system is beneficial. We have been running these benchmarks on an experimental test system and found some performance variances between the two sockets and the two nodes. The reason for this is not clear, although we hypothesize that this is due to the second socket having lower memory performance than the first socket because these differences are more significant for kernels that stress the memory subsystem.

We conclude that the SG2042 delivers impressive compute performance, especially when one considers the performance per dollar, and it is beneficial to build multi-socket systems based around this CPU. Indeed, for both compute and memory bound codes one is able to typically obtain improved performance by running across an additional socket and-so systems such as that provided by E4 Computer Engineering SPA are beneficial and the community should look to promote and adopt these.

\begin{acks}
We thank E4 Computer Engineering SPA for access to their dual-socket Sophon SG2042 test system that was used in this work, and especially Daniele Gregori, Elisabetta Boella and Fabrizio Magugliani for their advice and support. This work has been funded by the RISC-V EPSRC H\&ES ExCALIBUR testbed (EP/Y01412X/1).
\end{acks}

\bibliographystyle{ACM-Reference-Format}
\bibliography{sample-base}

\end{document}